\documentclass[a4paper]{jpconf}
\usepackage{graphicx}
\begin{document}

\newcommand{\R}{{\sf R\hspace*{-0.9ex}\rule{0.15ex}%
{1.5ex}\hspace*{0.9ex}}}
\def\bkR{{\rm I\kern-.17em R}}
\def\bkC{{\rm \kern.24em \vrule width.05em height1.4ex depth-.05ex
    \kern-.26em C}}  

\title{Brane Lorentz Symmetry from Lorentz Breaking in the
  Bulk\footnote{Based on a talk presented by O. Bertolami at
    D.I.C.E. 2006, Piombino, Italy.}} 

\author{O. Bertolami$^{1,2}$ and C. Carvalho$^{1,2}$}

\address{$^1$ Departamento de F\'{\i}sica, Instituto Superior
  T\'ecnico, Avenida Rovisco Pais 1, 1049-001 Lisboa, Portugal} 
\address{$^2$ Centro de F\'{\i}sica dos Plasmas, Instituto Superior
  T\'ecnico, Avenida Rovisco Pais 1, 1049-001 Lisboa, Portugal}  

\ead{orfeu@cosmos.ist.utl.pt, ccarvalho@ist.edu}

\begin{abstract}
We propose the mechanism of 
spontaneous symmetry breaking of a bulk vector field as a way 
to generate the selection of bulk dimensions invisible to the
standard model confined to the brane. By assigning a non-vanishing
vacuum value to the vector field, a direction is singled out in the
bulk vacuum, thus breaking the bulk Lorentz symmetry. We present the
condition for induced Lorentz symmetry on the brane, as
phenomenologically required.

\end{abstract}

\section{Introduction}

\vspace{0.2cm}

The prediction of extra dimensions, which the brane-confined standard
model is oblivious to, suggests that early enough in the history of
the Universe a dimensional selection would have taken place, implying
the breaking of the Lorentz symmetry at the level of the higher 
dimensional bulk spacetime
and the possible generation of a different geometry
along the directions orthogonal to the brane.

The breaking of the Lorentz symmetry can be realized by 
spontaneous breaking of a  
symmetry in the presence of a vector
field $B_{\mu}$ living in the bulk and subject to a potential $V(B^2).$ 
The acquisition by the bulk vector field $B_{\mu}$  of a
non-vanishing vacuum value  
and the consequent assignment to the bulk vacuum of an 
intrinsic direction determined by $\left< B_{\mu}\right>$ induce the
breaking of rotation invariance and thus of Lorentz symmetry in the
bulk.  
The brane, regarded as the locus of the observable universe embedded in
the higher dimensional bulk spacetime, would be expected to share a
subgroup of the symmetries of the bulk preserved in a manner akin to the
Goldstone mechanism.  
The overly tested Lorentz symmetry of the observable universe imposes that 
the brane must stand as a vacuum solution where the Lorentz invariance
would be a residual symmetry of the spontaneously broken  
symmetry \cite{BeCarvalho06}.

The possibility of violation of Lorentz invariance 
has nevertheless been widely discussed in the recent literature (see 
e.g. \cite{Kostelecky1}). Spontaneous breaking of Lorentz
symmetry may arise in the context of string/M-theory due to the existence of
non-trivial solutions in string field theory  \cite{Kostelecky2}, 
in loop quantum gravity \cite{Gambini} and  
in noncommutative field theories 
\cite{Carroll}, or via the spacetime variation of 
fundamental couplings \cite{Lehnert}.
Lorentz violation modifications of the dispersion
relations via five dimensional operators for fermions have also been
considered and constrained \cite{Bertolami2}.
Consequently, this putative breaking will have astrophysical
\cite{Sato,Bertolami1} implications, and in what concerns ultra-high 
energy cosmic rays, one can 
establish that Lorentz symmetry holds up to about
$2\times 10^{-25}$ \cite{Bertolami1}. The way to relate the breaking 
of Lorentz symmetry to gravity has been considered in Ref. \cite{Kostelecky3}, 
and solutions as well as implications have been discussed 
in Ref. \cite{Bertolami05}. 
In Ref. \cite{Bertolami3}, a connection between the cosmological constant 
and the violation of Lorentz invariance has been conjectured. 
For a general discussion on the relation between spacetime symmetries
and higher dimensions, the reader is referred to Ref.~\cite{Bertolami06}.

In this contribution we report on a recent study on 
how spontaneous Lorentz violation in the bulk has
repercussions on the brane and how it can be constrained.
We consider a bulk vector field coupled non-minimally to the graviton
which, upon acquiring a non-vanishing expectation
value in the vacuum, introduces spacetime anisotropies in the
gravitational field equations through the coupling with the graviton 
\cite{Kostelecky3,Bertolami05}. First we derive the bulk field equations 
and project them parallel and 
orthogonal to the brane. We then establish how to derive brane quantities
from bulk quantities by adopting Fermi normal coordinates with respect
to the directions on the brane and continuing into the bulk using the
Gauss normal prescription. Finally, we obtain the conditions for the 
Lorentz invariance to be a symmetry on the brane.

\section{Notation Definition}

\vspace{0.2cm}

We begin by parameterizing the world-sheet in terms of coordinates
$x^{A}=(t_b,{\bf x}_b)$ 
intrinsic to the brane. Using the chain rule, we may express the brane
tangent and normal unit vectors in terms of the bulk basis as follows:
\begin{eqnarray}
\hat e_A &=&{\partial \over \partial x^A}
=X_A^\mu{\partial \over \partial x^\mu} =X_A^\mu \hat e_\mu,\cr
\hat e_N &=&{\partial \over \partial n}
=N^\mu {\partial \over \partial x^\mu} =N^\mu \hat e_\mu,
\end{eqnarray}
with
\begin{eqnarray}
g_{\mu\nu}N^\mu N^\nu =1,\quad g_{\mu\nu}N^\mu X^\nu_A =0,
\end{eqnarray} 
where ${\bf g}$ is the bulk metric
\begin{eqnarray}
{\bf g}
&=&g_{\mu\nu}{\hat e}_\mu\otimes {\hat e}_\nu 
=g_{AB}~{\hat e}_A\otimes {\hat e}_B
+g_{AN}~{\hat e}_A\otimes {\hat e}_N 
+g_{NB}~{\hat e}_N\otimes {\hat e}_B
+g_{NN}~{\hat e}_N\otimes {\hat e}_N.
\end{eqnarray}
To obtain the metric induced on the brane we expand the bulk basis
vectors in terms of the coordinates intrinsic to the brane and keep the
doubly brane tangent components only. It follows that 
\begin{eqnarray}
g_{AB} =X^\mu _AX^\nu _B~g_{\mu\nu}
\end{eqnarray}
is the $(3+1)$-dimensional
metric induced on the brane by the $(4+1)$-dimensional bulk metric
$g_{\mu\nu}.$ 
The induced metric with upper indices is defined by the relation
\begin{eqnarray}
g_{AB}~g^{BC}=\delta _A{}^C.
\end{eqnarray}
It follows that we can write any bulk tensor field as a linear
combination of mutually orthogonal vectors on the brane, $\hat e_A,$ 
and a vector normal to the brane, $\hat e_N.$ We illustrate the
example of a vector $B_{\mu}$ and a tensor $T_{\mu\nu}$ bulk fields as follows
\begin{eqnarray}
{\bf B}
&=& B_{A}~{\hat e}_A +B_{N}~{\hat e}_N, \\
{\bf T} 
&=&T_{AB}~{\hat e}_A\otimes {\hat e}_B
+T_{AN}~{\hat e}_A\otimes {\hat e}_N 
+T_{NB}~{\hat e}_N\otimes {\hat e}_B
+T_{NN}~{\hat e}_N\otimes {\hat e}_N.
\end{eqnarray}
Derivative operators decompose similarly.
We write the derivative operator $\nabla$ as
\begin{eqnarray}
\nabla =(X^\mu_A +N^\mu)\nabla_\mu =\nabla_A +\nabla_N.
\end{eqnarray}
Fixing a point on the boundary, we introduce coordinates for the neighbourhood
choosing them to be Fermi normal.
All the Christoffel symbols of the metric on the boundary are
thus set to zero, although the partial derivatives do not in general
vanish. The non-vanishing connection coefficients are
\begin{eqnarray}
\nabla_A {\hat e}_B &=&-K_{AB}~{\hat e}_N, \cr
\nabla_A {\hat e}_N &=&+K_{AB}~{\hat e}_B, \cr
\nabla_N {\hat e}_A &=&+K_{AB}~{\hat e}_B, \cr
\nabla_N {\hat e}_N &=&0,
\end{eqnarray}
as determined by the Gaussian normal prescription for the continuation
of the coordinates off the boundary.
For the derivative operator $\nabla\nabla$ we find that
\begin{eqnarray}
\nabla\nabla&=& g^{\mu\nu}\nabla_\mu \nabla_\nu \cr
&=&g^{AB}\left[ (X^\mu_A\nabla_\mu)(X^\nu_B\nabla_\nu) 
-X^\mu_A(\nabla_\mu X^\nu_B)\nabla_\nu \right] 
+g^{NN}\left[ (N^\mu\nabla_\mu)(N^\nu\nabla_\nu) 
-N^\mu(\nabla_\mu N^\nu)\nabla_\nu \right] \cr
&=&g^{AB}\left[ \nabla_A \nabla_B +K_{AB}\nabla_N\right]
+               \nabla_N \nabla_N.
\end{eqnarray}
We can now decompose the Riemann tensor, $R_{\mu\nu\rho\sigma},$ along
the tangent and the normal directions to the surface of the brane as
follows
\begin{eqnarray}
R_{ABCD} &=& R_{ABCD}^{(ind)} +K_{AD}K_{BC} -K_{AC}K_{BD}, \qquad\\
R_{NBCD} &=& K_{BC;D} -K_{BD;C}, \\
R_{NBND} &=& K_{BC}K_{DC} -K_{BC,N},
\end{eqnarray}
from which we find the decomposition of the Einstein tensor,
$G_{\mu\nu},$ obtaining the Gauss-Codacci relations
\begin{eqnarray}
G_{AB} &=&G_{AB}^{(ind)} +2K_{AC}K_{CB} -K_{AB}K -K_{AB,N} 
-{1\over 2}g_{AB}\left( 3K_{CD}K_{DC} -K^2 -2K_{,N}\right),\quad \\
G_{AN} &=&K_{AB;B} -K_{;A}, \\
G_{NN} &=&{1\over 2}\left( -R^{(ind)} -K_{CD}K_{DC} +K^2\right).
\label{eqn:gc}
\end{eqnarray}

\section{Bulk Vector Field Coupled to Gravity}

\vspace{0.2cm}

In order to study the gravitational effects of the breaking of Lorentz
symmetry in a braneworld scenario, 
we consider a bulk vector field ${\bf B}$ with a non-minimal coupling 
to the graviton in a five-dimensional anti-de Sitter space. 
The Lagrangian density consists of
the Hilbert term, the cosmological constant term,
the kinetic and potential terms for 
${\bf B}$ and the ${\bf B}$--graviton interaction term, as follows
\begin{eqnarray}
{\cal L} 
={1\over {\kappa_{(5)}^2}}R -2\Lambda
+\xi B^{\mu}B^{\nu}R_{\mu\nu}
-{1\over 4}B_{\mu\nu}B^{\mu\nu} -V(B^\mu B_\mu \pm b^2 ),\quad
\end{eqnarray}
where $B_{\mu\nu} =\nabla_{\mu}B_{\nu} -\nabla_{\nu}B_{\mu}$ is the
tensor field associated with $B_{\mu}$ and $V$ is the potential which 
induces the spontaneous global symmetry breaking when the ${\bf B}$ field is
driven to the minimum at $B^\mu B_\mu \pm b^2 = 0 $, $b^2$ being  
a real positive constant. 
Here, $\kappa_{(5)}^2=8\pi G_{N}=M_{Pl}^3,$
$M_{Pl}$ is the five-dimensional Planck mass
and $\xi$ is a dimensionless coupling constant that we have
inserted to track the effect of the interaction\footnote{In 
Ref.~\cite{BeCarvalho06} we used $\xi =1,$ which enabled a 
simplification of the results without loss in generality of the 
purpose of the paper. Here, however, we shall keep $\xi$ free.}.
In the cosmological constant term
$\Lambda =\Lambda_{(5)} +\Lambda_{(4)}$ we have included both the bulk
vacuum value $\Lambda_{(5)}$ and that of the brane $\Lambda_{(4)},$
described by a brane tension $\sigma$ localized on the locus of the
brane, $\Lambda_{(4)}=\sigma\delta(N).$

By varying the action with respect to the metric, we 
obtain the Einstein equation
\begin{eqnarray}
{1\over \kappa_{(5)}^2}G_{\mu\nu} +\Lambda g_{\mu\nu}
-\xi L_{\mu\nu} -\xi \Sigma_{\mu\nu} ={1\over 2}T_{\mu\nu},
\end{eqnarray}
where  
\begin{eqnarray}
L_{\mu\nu} &=&{1\over 2}g_{\mu\nu}B^{\rho}B^{\sigma}R_{\rho\sigma} 
-\left( B_{\mu}B^{\rho}R_{\rho\nu} +R_{\mu\rho}B^{\rho}B_{\nu}\right),\qquad\\
\Sigma_{\mu\nu} &=& {1\over 2}\bigl[
\nabla_{\mu}\nabla_{\rho}(B_{\nu}B^{\rho}) 
+\nabla_{\nu}\nabla_{\rho}(B_{\mu}B^{\rho})\
-\nabla^2(B_{\mu}B_{\nu}) 
-g_{\mu\nu}\nabla_{\rho}\nabla_{\sigma}(B^{\rho}B^{\sigma})\bigr]
\end{eqnarray}
are the contributions from the interaction term and
\begin{eqnarray}
T_{\mu\nu} = 
B_{\mu\rho}B_{\nu}{}^{\rho} +4V^\prime B_{\mu}B_{\nu}
+g_{\mu\nu}\left[ -{1\over 4}B_{\rho\sigma}B^{\rho\sigma} -V\right] \quad
\end{eqnarray}
is the contribution from the vector field for the stress-energy tensor.
For the equation of motion for the vector field ${\bf B},$ obtained by
varying the action with respect to $B_{\mu},$ we find that
\begin{eqnarray}
\nabla^{\nu}
 \left( \nabla_{\nu}B_{\mu} -\nabla_{\mu}B_{\nu}\right)
-2V^{\prime}B_{\mu}
+2\xi B^{\nu}R_{\mu\nu} 
=0, \quad
\end{eqnarray}
where $V^{\prime} =dV/dB^2.$

We now proceed to project the equations parallel and orthogonal to
the surface of the brane. 
Following the prescription used in the derivation of the Gauss-Codacci
relations, we derive the components of the stress-energy tensor and of
the interaction terms.  
For the stress-energy tensor we find that
\begin{eqnarray}
T_{AB} &=&
B_{AC}B_{B}{}^{C} +B_{AN}B_{B}{}^{N} +4V^{\prime}B_{A}B_{B}
+g_{AB}\left[ -{1\over 4}\left( B_{CD}B^{CD} +2B_{CN}B^{CN}\right) 
-V\right], \cr
T_{AN} &=& B_{AC}B_{N}{}^{C} +4V^{\prime}B_{A}B_{N},\cr
T_{NN} &=& B_{NC}B_{N}{}^{C} +4V^{\prime}B_{N}B_{N}
+g_{NN}\left[ -{1\over 4}\left( B_{CD}B^{CD} +2B_{CN}B^{CN}\right) 
-V\right],
\end{eqnarray}
and for the interaction source terms that
\begin{eqnarray}
L_{AB}&=&
{1\over 2}g_{AB}\Big[
B^{C}B^{D}\left( R^{(ind)}_{CD} +2K_{CE}K_{ED} -K_{CD}K -K_{CD,N}\right)\cr
&&\quad\quad{}+2B^{C}B^{N}\left( K_{EC;E} -K_{;C}\right)
+B^{N}B^{N}\left( K_{CD}K_{DC} -K_{,N}\right)\Big]\cr
&&{}-B_{A}\left[
B^{C}\left( R^{(ind)}_{CB} +2K_{CE}K_{EB} -K_{CB}K -K_{CB,N}\right)
+B^{N}\left( K_{EB;E} -K_{;B}\right)\right]\cr
&&-\left[ 
\left( R^{(ind)}_{AC} +2K_{AE}K_{EC} -K_{AC}K -K_{AC,N}\right)B^{C}
+\left( K_{EA;E} -K_{;A}\right)B^{N}\right]B_{B},
\cr
L_{AN}&=&
{}-B_{A}\left[ 
B^{C}\left( K_{EC;E} -K_{;C}\right) 
+B^{N}\left( K_{EH}K_{EH} -K_{,N}\right)\right] \cr
&&{}-\left[ 
\left( R^{(ind)}_{AC} +2K_{AE}K_{EC} -K_{AC}K -K_{AC,N}\right)B^{C}
+\left( K_{EA;E} -K_{;A}\right)B^{N}\right]B_{N},\cr
L_{NN}&=&
{}-2B_{N}\left[ B^{C}\left( K_{EC;E} -K_{;C}\right) 
 +B^{N}\left( K_{EH}K_{EH} -K_{,N}\right)\right],
\end{eqnarray}
and 
\begin{eqnarray}
\Sigma_{AB}&=&
{1\over 2}\Big[
\nabla_{A}\nabla_{C}(B_{B}B_{C}) +\nabla_{A}\nabla_{N}(B_{B}B_{N})
+\nabla_{B}\nabla_{C}(B_{A}B_{C}) +\nabla_{B}\nabla_{N}(B_{A}B_{N}) \cr
&&{}-\left( \nabla_{C}\nabla_{C} +\nabla_{N}\nabla_{N}\right)(B_{A}B_{B}),
\cr
&&{}-g_{AB}\left[ 
\nabla_{C}\nabla_{D}(B_{C}B_{D}) +\nabla_{C}\nabla_{N}(B_{D}B_{N})
+\nabla_{N}\nabla_{C}(B_{N}B_{C}) +\nabla_{N}\nabla_{N}(B_{N}B_{N})\right] 
\cr
&&{}+2K_{AB}\left( \nabla_{C}(B_{N}B_{C}) +\nabla_{N}(B_{N}B_{N})\right)
+K\left( \nabla_{A}(B_{B}B_{N}) +\nabla_{B}(B_{A}B_{N})\right) \cr
&&{}-2K_{AC}\left( \nabla_{C}(B_{B}B_{N}) -\nabla_{N}(B_{B}B_{C})\right)
-2K_{BC}\left( \nabla_{C}(B_{A}B_{N}) -\nabla_{N}(B_{A}B_{C})\right)
-K\nabla_{N}(B_{A}B_{B}) \cr
&&{}-g_{AB}\left[ 
2K\nabla_{C}(B_{C}B_{N}) -2K_{CD}\nabla_{C}(B_{D}B_{N})
+K\nabla_{N}(B_{N}B_{N}) -K_{CD}\nabla_{N}(B_{C}B_{D})\right] \cr
&&{}+(\nabla_{A}K)B_{B}B_{N} +(\nabla_{B}K)B_{A}B_{N} \cr
&&{}-(\nabla_{C}K_{CA})B_{N}B_{B} +(\nabla_{N}K_{CA})B_{C}B_{B}
-(\nabla_{C}K_{CB})B_{A}B_{N} +(\nabla_{N}K_{CB})B_{A}B_{C}\cr
&&{}-g_{AB}\left[
(\nabla_{C}K)B_{C}B_{N} -(\nabla_{N}K_{CD})B_{D}B_{C} 
+(\nabla_{N}K)B_{N}B_{N}\right] \cr
&&{}+2K_{AB}( -K_{CD}B_{C}B_{D} +KB_{N}B_{N})\cr
&&{}-2K_{AC}K_{BC}B_{N}B_{N} -2K_{AC}K_{BD}B_{C}B_{D}
+K( K_{AC}B_{C}B_{B} +K_{BC}B_{A}B_{C})\cr
&&{}-g_{AB}\left[
(2K_{CE}K_{ED} -KK_{CD})B_{C}B_{D} -(2K_{CD}K_{DC} -K^2)B_{N}B_{N}
\right]
\Big]\cr
\cr
\Sigma_{AN}&=&
{1\over 2}\Big[
\nabla_{A}\nabla_{C}(B_{N}B_{C}) +\nabla_{A}\nabla_{N}(B_{N}B_{N})
+\nabla_{N}\nabla_{C}(B_{A}B_{C}) 
-\nabla_{C}\nabla_{C} 
(B_{A}B_{N})
\cr
&&{}-2K_{AC}\left[ \nabla_{D}(B_{D}B_{C}) +\nabla_{C}(B_{N}B_{N})\right]
-K_{CD}\nabla_{A}(B_{C}B_{D}) +K\nabla_{A}(B_{N}B_{N}) 
+K_{AC}\nabla_{N}(B_{C}B_{N}) \cr
&&{}-(\nabla_{A}K_{CD})B_{D}B_{C} 
+(\nabla_{A}K)B_{N}B_{N} 
+(\nabla_{N}K)B_{A}B_{N} +(\nabla_{C}K_{CD})B_{A}B_{D} \cr
&&{}-(\nabla_{C}K_{CA})B_{N}B_{N} +(\nabla_{N}K_{AC})B_{C}B_{N} \cr
&&{}-K_{AC}KB_{C}B_{N} -K_{CD}K_{DC}B_{A}B_{N}
\Big],\cr
\cr
\Sigma_{NN}&=&
{1\over 2}\Big[
\nabla_{N}\nabla_{C}(B_{N}B_{C}) 
-\nabla_{C}\nabla_{D}(B_{C}B_{D}) -\nabla_{C}\nabla_{N}(B_{C}B_{N}) 
-\nabla_{C}\nabla_{C} 
(B_{N}B_{N})
\cr
&&{}-K_{CD}\nabla_{N}(B_{D}B_{C})
-2K\nabla_{C}(B_{N}B_{C}) +2K_{CD}\nabla_{C}(B_{N}B_{D}) \cr
&&{}-(\nabla_{N}K_{CD})B_{D}B_{C} 
+(\nabla_{N}K)B_{N}B_{N} 
+2(\nabla_{C}K_{CD})B_{D}B_{N} -(\nabla_{C}K)B_{C}B_{N} \cr
&&{}+KK_{CD}B_{D}B_{C} -K^2B_{N}B_{N}
\Big].
\end{eqnarray}
The equation of motion for the vector field ${\bf B}$
decomposes similarly as follows, respectively parallel 
\begin{eqnarray}
&&\nabla_{C}\left( \nabla_{C}B_{A} -\nabla_{A}B_{C}\right) 
+\nabla_{N}\left( \nabla_{N}B_{A} -\nabla_{A}B_{N}\right)\cr
&+&2K_{AC}\left( \nabla_{C}B_{N} -\nabla_{N}B_{C}\right)
+K\left( \nabla_{N}B_{A} -\nabla_{A}B_{N}\right)
-2V^{\prime}B_{A}\cr 
&+&2\xi \Bigl[ 
B_{C}\left( R^{(ind)}_{AC} +2K_{AD}K_{DC} -K_{AC}K -K_{AC,N}\right)
+B_{N}\left( K_{AC;C} -K_{;A}\right) \Bigr]
=0,
\label{eqn:B_A}
\end{eqnarray}
and orthogonal to the brane
\begin{eqnarray}
&&\nabla_{C}\left( \nabla_{C}B_{N} -\nabla_{N}B_{C}\right) 
-2V^{\prime}B_{N}\cr 
&+&2\xi \left[ 
B_{C}\left( K_{CD;D} -K_{;C}\right)
+B_{N}\left( K_{CD}K_{CD} -K_{;N}\right) \right] = 0.
\label{eqn:B_N}
\end{eqnarray}

Next we proceed to derive the induced equations of motion for both
the metric and the vector field in terms of quantities measured on the
brane. The induced equations on the brane are the $(AB)$ projected
components after the singular terms across the brane are subtracted
out by the substitution of the matching conditions.
Considering the brane as a $Z_2$-symmetric shell of
thickness $2\delta$ in the limit $\delta \to 0,$ 
derivatives of quantities discontinuous across the brane generate
singular distributions on the brane. Integration of these terms in the
coordinate normal to the brane
relates the induced geometry 
with the localization of the induced stress-energy 
in the form of matching conditions.
From the $Z_{2}$ symmetry it follows that 
$B_{A}(-\delta) =+B_{A}(+\delta)$ 
but that $B_{N}(-\delta) =-B_{N}(+\delta),$ and consequently that 
$(\nabla_{N}B_{A})(-\delta) =-(\nabla_{N}B_{A})(+\delta)$  and 
$(\nabla_{N}B_{N})(-\delta) =+(\nabla_{N}B_{N})(+\delta).$
Moreover,
$g_{AB}(N=-\delta) =+g_{AB}(N=+\delta)$ implies that 
$K_{AB}(N=-\delta) =-K_{AB}(N=+\delta).$
First we consider the Einstein equations and then the equations of
motion for ${\bf B}$ which, due to the coupling of ${\bf B}$ to
gravity, will also be used as complementary conditions for the dynamics of
the metric on the brane.

Combining the Gauss-Codacci relations with the projections of the
stress-energy tensor and the interaction source terms, we integrate
the $(AB)$ component of the Einstein equation in the 
coordinate normal to the brane to obtain the matching conditions for
the extrinsic curvature across the brane, i.e. the Israel matching
conditions.
We find 
that
\begin{eqnarray}
&&{1\over \kappa _{(5)}^2}
\left[ -K_{AB} +g_{AB}K\right]
=
-g_{AB}\sigma
\cr
&+&{\xi \over 2}\Big[
\nabla_{A}(B_{B}B_{N}) +\nabla_{B}(B_{A}B_{N})
-\nabla_{N}(B_{A}B_{B}) \cr
&&{}+4( B_{A}B_{C}K_{CB} +K_{AC}B_{C}B_{B})
-2K_{AB}B_{N}B_{N}\cr
&&{}+g_{AB}\left( 
-2\nabla_{C}(B_{C}B_{N}) -\nabla_{N}(B_{N}B_{N}) 
+K_{CD}B_{C}B_{D} -KB_{N}B_{N}\right)
\Big].
\label{eqn:imc}
\end{eqnarray}
These provide boundary conditions for ten of the fifteen degrees of
freedom. Five additional boundary conditions are provided by the
junction conditions for the $(AN)$ and 
$(NN)$ components of the projection of the Einstein equations.
From inspection of the $(AN)$ component, we note that 
\begin{eqnarray}
G_{AN} &=& K_{AB;B} -K_{;A} 
= -\nabla_{B}\left( \int ^{+\delta}_{-\delta}
dN~G_{AB}\right)
=-\kappa_{(5)}^2 \nabla_{B}{\cal T}_{AB} =0
\end{eqnarray}
which vanishes due to conservation of the induced stress-energy tensor
${\cal T}_{AB}$ on the brane.
From integration of the $(NN)$ component in the normal direction to the brane,
we find the following junction condition
\begin{eqnarray}
\nabla_{C}(B_{C}B_{N}) +3KB_{N}B_{N} -K_{CD}B_{C}B_{D} =0, 
\label{eqn:G_NN:junction}
\end{eqnarray}
which we substitute back in, obtaining
\begin{eqnarray}
&&{1\over \kappa_{(5)}^2}{1\over 2}
\left( -R^{(ind)} -K_{CD}K_{CD} +K^2\right)
+\Lambda_{(5)}\cr
&=&{1\over 2}
\left[ -{1\over 4}B_{CD}B_{CD} 
+4V^{\prime}B_{N}B_{N} -V\right] \cr
&+&{\xi\over 2}\Bigl[
-\nabla_{C}\nabla_{D}(B_{C}B_{D})
-\nabla_{C}\nabla_{C}(B_{N}B_{N})
+{12\over {2\xi -1}}B_{N}\left( 
\nabla_{C}\nabla_{C}B_{N} -2V^{\prime}B_{N}\right)\cr
&&{}+2\left( K_{CD} -g_{CD}K\right)\nabla_{C}(B_{D}B_{N})
+2K_{CD}B_{D}(\nabla_{C}B_{N})
+K_{CD;C}B_{D}B_{N}\cr
&&{}
+\left( 7K_{CD}K_{CD} -K^2\right)B_{N}B_{N}
+\left( (6\xi +1)K_{CE}K_{ED} +KK_{CD}\right)B_{C}B_{D}\Bigr].
\label{eqn:G_NN:induced}
\end{eqnarray}
However, the Israel matching conditions also contain terms which depend on the
prescription for the continuation of ${\bf B}$ 
out of the brane and into the bulk, 
namely $\nabla_{N}B_{A}$ and $\nabla_{N}B_{N}$. The five
additional boundary conditions required are those for the vector field
{\bf B.}
In Ref.~\cite{BuCarvalho05} 
the boundary conditions for bulk fields were derived subject to 
the condition that 
modes emitted by the brane into the bulk  
do not violate the gauge defined in the bulk.
Here, however,  
we integrate the (A) and (N) components of the equation of
motion for ${\bf B},$ Eq.~(\ref{eqn:B_A}) and Eq.~(\ref{eqn:B_N})
respectively, to find the corresponding junction condition for $B_{A}$
and for $B_{N}$ across the brane. 
From Eq.~(\ref{eqn:B_A}) we have that
\begin{eqnarray}
\int ^{+\delta}_{-\delta} dN \bigl[ 
\nabla_{N}\left(
\nabla_{N}B_{A} -\nabla_{A}B_{N}\right)
-2\xi K_{AC}(\nabla_{N}B_{C}) 
-2\xi B_{C}K_{AC,N}\bigr] 
=0.
\end{eqnarray}
If $\delta$ is sufficiently small, the difference between $K_{AB;N}$
and $K_{AB,N}$ is negligibly small. It follows that, in the limit where
$\delta \to 0,$ we can assume that $\nabla_{N}\approx \partial_{N}$. It 
then follows that
\begin{eqnarray}
\nabla_{N}B_{A} -\nabla_{A}B_{N} 
-2\xi K_{AC}B_{C}
=0.
\label{eqn:B_A:junction}
\end{eqnarray}
Similarly, from Eq.~(\ref{eqn:B_N}) we find that
\begin{eqnarray}
\int ^{+\delta}_{-\delta} dN \left[ 
-\nabla_{N}\nabla_{C}B_{C} 
-(2\xi -1) \nabla_{N}(K B_{N})\right]=0,
\end{eqnarray}
which becomes
\begin{eqnarray}
\nabla_{C}B_{C} +(2\xi -1) K B_{N} 
=0.
\label{eqn:B_N:junction}
\end{eqnarray}
The junction conditions Eq.~(\ref{eqn:B_A:junction}) and 
Eq.~(\ref{eqn:B_N:junction}) offer the required $(4+1)$ boundary
conditions respectively for $B_{A}$ and $B_{N}$ on the brane. 
Substituting the junction condition for $B_{A}$
back in Eq.~(\ref{eqn:B_A}) and using the result from $G_{AN}=0,$
we find for the induced equation of motion for $B_{A}$ on the brane that
\begin{eqnarray}
\nabla_{C}\left( \nabla_{C}B_{A} -\nabla_{A}B_{C}\right) 
+2\xi K_{AC}(\nabla_{C}B_{N})
-2V^{\prime}B_{A} 
+2\xi 
B_{C}\left( R^{(ind)}_{AC} +2\xi K_{AD}K_{DC}\right)
=0.
\label{eqn:B_A:induced}
\end{eqnarray}
Similarly, substituting the junction condition for $B_{N}$ 
back in Eq.~(\ref{eqn:B_N}) we obtain
\begin{eqnarray}
\nabla_{C}\nabla_{C}B_{N} 
-2V^{\prime}B_{N} 
+(2\xi -1) \left[
K(\nabla_{N}B_{N})
+B_{N}K_{CD}K_{CD} \right]
=0.
\label{eqn:B_N:induced}
\end{eqnarray}
Thus, Eq.~(\ref{eqn:B_A:junction}) provides the value at the boundary
for $\nabla_{N}B_{A}$ whereas Eq.~(\ref{eqn:B_N:induced}) provides that
for $\nabla_{N}B_{N}$.
Using the results derived above in the Israel matching conditions we find that
\begin{eqnarray}
&&{1\over \kappa _{(5)}^2}
\left[ -K_{AB} +g_{AB}K\right]
=-g_{AB}~\sigma \cr
&+&{\xi\over 2}\Big[
(\nabla_{A}B_{B})B_{N} +(\nabla_{B}B_{A})B_{N}
\Big]\cr
&+& \xi B_{A}B_{C}K_{CB} +\xi K_{AC}B_{C}B_{B} -K_{AB}B_{N}B_{N}\cr
&+&\xi g_{AB}\biggl[ 
-\nabla_{C}(B_{C}B_{N}) 
+{1\over 2}K_{CD}B_{C}B_{D} -{1\over 2}KB_{N}B_{N}\cr
&&{}+{1\over {2\xi -1}}{B_{N}\over K} \bigl(
\nabla_{C}\nabla_{C} B_{N}
-2V^{\prime}B_{N}
+(2\xi -1)K_{CD}K_{CD}B_{N}\bigr)
\biggr].
\end{eqnarray}
which provide a second order equation for the trace of
the extrinsic curvature, $K.$
Finally, using Eq.~(\ref{eqn:G_NN:induced})
in the $(AB)$ Einstein equation, we find for the 
Einstein equation induced on the brane that
\begin{eqnarray}
&&{1\over \kappa_{(5)} ^2}\left[ 
G_{AB}^{(ind)} +2K_{AC}K_{BC} -K_{AB}K 
+{1\over 2}g_{AB}\left( -R^{(ind)} -4K_{CD}K_{CD} +2K^2\right)\right] 
+2g_{AB}\Lambda_{(5)}\cr
&+&{}{1\over 2}\left[ 
-B_{AC}B_{BC} -4V^{\prime}B_{A}B_{B} 
+g_{AB}\left( {1\over 4}B_{CD}B_{CD} 
+4V^{\prime}B_{N}B_{N}
+2V\right)\right]\cr
&=&{\xi\over 2}\biggl[
\nabla_{A}\nabla_{C}(B_{B}B_{C})
+\nabla_{B}\nabla_{C}(B_{A}B_{C})
-\nabla_{C}\nabla_{C}(B_{A}B_{B})\cr
&&{}
-2K_{AC}\nabla_{C}( B_{B}B_{N}) 
-2K_{AC}(B_{B}\nabla_{C}B_{N} +B_{C}\nabla_{B}B_{N})\cr
&&{}
-2K_{BC}\nabla_{C}( B_{A}B_{N}) 
-2K_{BC}(B_{A}\nabla_{C}B_{N} +B_{C}\nabla_{A}B_{N})\cr
&&{}
-{8\over {2\xi -1}}{B_{N}\over K} K_{AB}
 ( \nabla_{C}\nabla_{C}B_{N} -2V^{\prime}B_{N} +(2\xi -1)K_{CD}K_{CD}B_{N})
+KB_{N}(\nabla_{A}B_{B} +\nabla_{B}B_{A}) \cr
&&{}
-2B_{A}B_{C}\left( R^{(ind)}_{CB} +2K_{CD}K_{BD} +(\xi -1)K_{CB}K\right)\cr
&&{}
-2B_{B}B_{C}\left( R^{(ind)}_{CA} +2K_{AD}K_{CD} +(\xi -1)K_{AC}K\right)\cr
&&{}+(K_{AC;B} +K_{BC;A} -2K_{AB;C})B_{N}B_{N} 
-4K_{AB}KB_{N}B_{N}\cr
&&{}
+( K_{AC}B_{B} +K_{BC}B_{A})( -5\xi K_{DC}B_{D} +KB_{C})
-(4\xi +2)K_{AC}K_{BD}B_{C}B_{D} 
\biggr]\cr
&+&{\xi\over 2}g_{AB}\biggl[
-2\nabla_{C}\nabla_{D}(B_{C}B_{D})
-\nabla_{C}\nabla_{C}(B_{N}B_{N})
+{12\over {2\xi -1}}B_{N}\left( 
\nabla_{C}\nabla_{C}B_{N} -2V^{\prime}B_{N}\right)
\cr
&&{}
+4(K_{CD} -g_{CD}K)\nabla_{D}(B_{C}B_{N})
+4K_{CD}B_{D}(\nabla_{C}B_{N}) \cr
&&{}
+B_{C}B_{D}R^{(ind)}_{CD}
+( 9K_{CD}K_{CD} -2K^2)B_{N}B_{N} 
+( 14\xi K_{CE}K_{DE} +KK_{CD})B_{C}B_{D} 
\biggr].
\end{eqnarray}
The results obtained above show both the coupling of the bulk to the
brane and the coupling of the vector field ${\bf B}$ to the geometry
of the spacetime. The first is manifested in the dependence on normal
components in the induced equations; the latter is manifested in the
presence of terms of the form $(R_{AB}B_{C}B_{D}).$ Terms of the form
$(K_{AB}B_{N})$ illustrate both couplings, where $B_{N}$ relates with
$K$ and $B_{A}$ by Eq.~(\ref{eqn:B_N:junction}). The directional
dependence on the $N$ direction is encapsulated in the extrinsic
curvature. 
In the penultimate line we can substitute the Israel matching condition
found above. 
However, 
the derivatives of the extrinsic curvature
along directions parallel to the brane which appear in the ninth line
are not reducible to quantities intrinsic to the brane.

\section{Bulk Vector Field with a Non-vanishing
Vacuum Expectation Value}

\vspace{0.2cm}

In this section we particularize the formalism developed above for
the case when the bulk vector field ${\bf B}$ acquires a non-vanishing
vacuum expectation value by spontaneous symmetry breaking. 
The vacuum value generates the breaking of the Lorentz
symmetry by selecting the direction orthogonal to the plane of the
brane. This selection can be achieved by 
assigning to the parallel or the orthogonal components of the bulk
vector field a non-vanishing vacuum expectation value. 
The minimum of the potential occupied by the vacuum value is
assumed to be also a zero of the potential.

\subsection{$\left<B_{A}\right> \not=0$ and $\left<B_{N}\right> =0$}
\label{1}

\vspace{0.2cm}

Here we consider the case when 
the parallel component of the vector field with respect to
the brane acquires a non-vanishing
expectation value, $\left<B_{A}\right>\not=0$, 
whereas the expectation
value of the normal component is chosen 
to vanish on the brane, $\left<B_{N}\right> =0$. 
The junction conditions from the equations for $B_{A},$ $B_{N},$
$G_{NN}$ and $G_{AB}$ reduce respectively to 
\begin{eqnarray}
\nabla_{N}\left<B_{A}\right> -2\xi K_{AC}\left<B_{C}\right> &=&0,\\
\nabla_{C}\left<B_{C}\right>&=&0,\\
K_{CD}\left<B_{C}\right>\left<B_{D}\right> &=&0,\\
{1\over \kappa_{(5)}^2}\left[ -K_{AB} +g_{AB}K\right]
&=&-g_{AB}~\sigma
+\xi \left<B_{A}\right>\left<B_{C}\right>K_{CB}
+\xi \left<B_{B}\right>\left<B_{C}\right>K_{AC},
\end{eqnarray}
and the induced equations of motion become
\begin{eqnarray}
\nabla_{C}\left( \nabla_{C}\left<B_{A}\right> 
-\nabla_{A}\left<B_{C}\right>\right)
+2\xi \left<B_{C}\right>\left( R_{AC}^{(ind)} +2\xi K_{AD}K_{DC}\right)=0
\end{eqnarray}
for $B_{A},$
\begin{eqnarray}
&&{1\over \kappa_{(5)}^2}{1\over 2}\left( 
-R^{(ind)} -K_{CD}K_{CD} +K^2\right)
+\Lambda_{(5)}\cr
&=&{1\over 2}\left[
-{1\over 4}\left< B_{CD}\right>\left< B_{CD}\right>\right]
+{\xi\over 2}\left[
-\nabla_{C}\nabla_{D}(\left< B_{C}\right>\left< B_{D}\right>)
+(6\xi +1)K_{CE}K_{ED}\left< B_{C}\right>\left< B_{D}\right>
\right]
\end{eqnarray}
for $G_{NN}$ and finally
\begin{eqnarray}
&&{1\over \kappa_{(5)} ^2}\biggl[ 
G_{AB}^{(ind)} +2K_{AC}K_{BC} 
-\left( {1\over 2} +\xi -1\right)K_{AB}K \cr
&&+{1\over 2}g_{AB}\left( 
R^{(ind)} -2K_{CD}K_{CD} 
-\left( 1 -2(\xi -1)\right)K^2\right)\biggr] 
+g_{AB}\Lambda_{(5)}
-{1\over 2}\left< B_{AC}\right>\left< B_{BC}\right> 
\cr
&=&
{1\over 2}\left( {5\over 4} +{1\over \xi}\right)\left[
 \left< B_{A}\right>\nabla_{C}\left( 
 \nabla_{C}\left< B_{B}\right>
 -\nabla_{B}\left< B_{C}\right>\right)
+\left< B_{B}\right>\nabla_{C}\left( 
 \nabla_{C}\left< B_{A}\right>
 -\nabla_{A}\left< B_{C}\right>\right)
\right]\cr
&+&{\xi\over 2}\biggl[
\nabla_{A}\nabla_{C}(\left< B_{B}\right>\left< B_{C}\right>)
+\nabla_{B}\nabla_{C}(\left< B_{A}\right>\left< B_{C}\right>)
-\nabla_{C}\nabla_{C}(\left< B_{A}\right>\left< B_{B}\right>)\cr
&&{}
+\left( {5\over 2} -2 +{2\over \xi}\right)\left( 
\left<B_{A}\right>\left<B_{C}\right>R^{(ind)}_{CB}
+\left<B_{B}\right>\left<B_{C}\right>R^{(ind)}_{AC}\right)
-(4\xi +2)K_{AC}K_{BD}\left< B_{C}\right>\left< B_{D}\right>
\biggr]\cr
&+&{1\over 2}(1 -2(\xi -1))g_{AB}K\sigma
+{\xi\over 2}g_{AB}\biggl[
\left< B_{C}\right>\left< B_{D}\right>R^{(ind)}_{CD}
+2(\xi -1)K_{CE}K_{ED}\left< B_{C}\right>\left< B_{D}\right>\biggr]
\label{eqn:Einsteinvev1}
\end{eqnarray}
from $G_{AB},$ where we used also the previous results, namely the
$G_{NN}$ equation, the Israel matching condition and the $B_{A}$ equation.

Imposing the condition for the covariant conservation of
the vacuum expectation value of the field ${\bf B},$ 
$\nabla_{A}\left<B_{C}\right> = 0$ \cite{Kostelecky3, Bertolami05},
it follows that 
$\left<B_{AC}\right>=\nabla_{A}\left<B_{C}\right>
-\nabla_{C}\left<B_{A}\right> =0,$
which enables us to further simplify Eq.~(\ref{eqn:Einsteinvev1}):
\begin{eqnarray} 
&&{1\over \kappa_{(5)} ^2}\biggl[ 
G_{AB}^{(ind)} +2K_{AC}K_{BC} 
-\left( {1\over 2} +\xi -1\right)K_{AB}K \cr
&&+{1\over 2}g_{AB}\left( 
R^{(ind)} -2K_{CD}K_{CD} 
-\left( 1 -2(\xi -1)\right)K^2\right)\biggr] 
+g_{AB}\Lambda_{(5)}
\cr
&=&
{\xi\over 2}\biggl[
\left( {5\over 2} -2 +{2\over \xi}\right)\left( 
\left<B_{A}\right>\left<B_{C}\right>R^{(ind)}_{CB}
+\left<B_{B}\right>\left<B_{C}\right>R^{(ind)}_{AC}\right)
-(4\xi +2)K_{AC}K_{BD}\left< B_{C}\right>\left< B_{D}\right>
\biggr]\cr
&+&{1\over 2}(1 -2(\xi -1))g_{AB}K\sigma
+{\xi\over 2}g_{AB}\biggl[
\left< B_{C}\right>\left< B_{D}\right>R^{(ind)}_{CD}
+2(\xi -1)K_{CE}K_{ED}\left< B_{C}\right>\left< B_{D}\right>\biggr].
\label{eqn:Einsteinvev2}
\end{eqnarray}
Hence, in order to obtain a vanishing cosmological constant
and ensure that Lorentz invariance holds on the brane, we must impose
respectively that
\begin{equation}
\Lambda_{(5)} = {1\over 2}(1 -2(\xi -1))K\sigma
\label{eqn:vanishingcc}
\end{equation} 
and that
\begin{eqnarray}
&&{1\over \kappa_{(5)} ^2}\biggl[
2K_{AC}K_{BC} 
-\left( {1\over 2} +\xi -1\right)K_{AB}K \cr
&&+{1\over 2}g_{AB}\left( 
R^{(ind)} -2K_{CD}K_{CD} 
-\left( 1 -2(\xi -1)\right)K^2\right)
\biggr]
\cr
&=& 
{\xi\over 2}\biggl[
\left( {5\over 2} -2 +{2\over \xi}\right)\left( 
\left<B_{A}\right>\left<B_{C}\right>R^{(ind)}_{CB}
+\left<B_{B}\right>\left<B_{C}\right>R^{(ind)}_{AC}\right)
-(4\xi +2)K_{AC}K_{BD}\left< B_{C}\right>\left< B_{D}\right>
\biggr]\cr
&+&{\xi\over 2}g_{AB}\biggl[
\left< B_{C}\right>\left< B_{D}\right>R^{(ind)}_{CD}
+2(\xi -1)K_{CE}K_{ED}\left< B_{C}\right>\left< B_{D}\right>\biggr]
\label{eqn:Lorentzbrane},
\end{eqnarray}
which for $\xi =1$ reduce to the results presented in
Ref.~\cite{BeCarvalho06}. 
We observe that there is close relation between the vanishing of the 
cosmological constant and the maintenance of the Lorentz invariance on the 
brane. 
These conditions are enforced so that the higher dimensional 
signatures encapsulated in the induced geometry of the brane cancel
the Lorentz symmetry breaking inevitably induced on the brane, thus
reproducing the observed geometry. The first condition, 
Eq.~(\ref{eqn:vanishingcc}), can be modified to account for any
non-vanishing value for the cosmological constant, as appears to be 
suggested by the recent Wilkinson Microwave Anisotropy Probe 
data, by defining the observed
cosmological constant $\Lambda$ such that 
$\Lambda_{(5)}= \Lambda +\tilde \Lambda_{(5)}.$ 
A much more elaborate fine-tuning, however, is required 
for the Lorentz symmetry to be observed on the brane, as described by 
the condition in Eq.~(\ref{eqn:Lorentzbrane}).
To our knowledge this is 
a new feature in braneworld models, as in most models Lorentz
invariance is a symmetry  
shared by both the bulk and the brane. Notice that a connection between the 
cosmological constant and Lorentz symmetry has been conjectured, on different 
grounds, long ago \cite{Bertolami3}.

\subsection{$\left<B_{A}\right> =0$ and $\left<B_{N}\right> \not=0$}
\label{2}

\vspace{0.2cm}

Choosing instead $\left<B_{A}\right> =0$ and 
$\left<B_{N}\right> \not=0$ 
we would also expect to violate Lorentz symmetry on the brane. 
However, for $K\not=0$ the boundary conditions imply that we must have
$\left<B_{N}\right> =0$ and thus rendering the vacuum Lorentz symmetric. 
If, on the other hand, we allow $K=0,$ then the Israel matching conditions 
yield that $\sigma =0$, 
rendering the brane inexistent, with $\left< B_{N}\right>$ being but 
the five dimensional gravitational constant 
$\left< B_{N}\right>=\pm 1/\kappa_{(5)}.$

\subsection{$\left<B_{A}\right> \not=0$ and $\left<B_{N}\right> \not=0$}
\label{3}

\vspace{0.2cm}

If we consider the general case, with both $B_{A}$ and $B_{N}$ acquiring
different, constant non-vanishing vacuum expectation values along the
directions parallel to the brane, 
i.e. $\left<B_{A}\right> \not=0$ and $\left<B_{N}\right> \not=0$
and such that  
$\nabla_{B}\left< B_{A}\right>=\nabla_{B}\left< B_{N}\right> =0$, 
we find that for $K\not=0$ we
must have that $\left<B_{N}\right> =0,$ thus obtaining the same boundary
conditions as those found in Subsection \ref{1}. 
Should we allow $K=0,$ then we find, as in Subsection \ref{2},
that $\sigma =0$ and that $\kappa_{(5)}$ is defined in terms
of both $\left< B_{A}\right>$ and $\left< B_{N}\right>$ according to 
$1/\kappa_{(5)} ^2 =
-\lambda [
\left< B_{A}\right>\left< B_{C}\right>K_{CB}
+\left< B_{B}\right>\left< B_{C}\right>K_{CA}]/K_{AB}
+\left< B_{N}\right>\left< B_{N}\right>$.  

\section{Discussion and Conclusions}

\vspace{0.2cm}

In this contribution we analysed the spontaneous symmetry breaking of
Lorentz invariance in the bulk and its effect on the brane. 
For this purpose, we considered a bulk vector field subject to a
potential which endows the field with a non-vanishing  
vacuum expectation value, thus allowing for the spontaneous breaking
of the Lorentz symmetry in the bulk. This bulk vector field is   
directly coupled to the Ricci tensor so that, after the breaking of
Lorentz invariance, the loss of  
this symmetry is transmitted to the gravitational sector.
We assign a non-vanishing vacuum expectation value first separately 
to the parallel and orthogonal components of the vector field, finding
then that the case where both components attained non-vanishing vacuum
expectation values reduced to the previous two cases. 
The complex interplay between matching conditions and the Lorentz symmetry 
breaking terms was examined. We found that Lorentz invariance 
on the brane can be made exact via the dynamics of the graviton, vector field
and the extrinsic curvature of the surface of the 
brane. As a consequence of the exact reproduction of Lorentz symmetry on
the brane, we found a condition for the matching of the observed
cosmological constant in four dimensions.  
This tuning does not follow from a dynamical mechanism but 
is instead imposed by phenomenological reasons only. 
From this point of view, both the value of the cosmological constant
and the induced brane Lorentz symmetry seem to be a consequence of a
complex fine tuning. We shall examine    
further implications of this mechanism in a forthcoming publication
where we will also discuss the inclusion of a bulk scalar field
\cite{BeCarvalho07}.

\vfill
\newpage

\centerline{\bf {Acknowledgments}}

\vskip 0.2cm

\noindent 
C. C. thanks the Portuguese Agency, Funda\c c\~ao para a Ci\^encia e a 
Tecnologia (FCT), for financial support under the
fellowship /BPD/18236/2004. C. C. thanks Martin Bucher, Georgios
Kofinas and Rodrigo Olea for useful discussions. The work of O.B. 
is partially supported by the FCT project POCI/FIS/56093/2004.



\end{document}